\documentclass[preprintnumbers,showpacs,amsmath,amssymb,floatfix,prd,onecolumn,superscriptaddress,nofootinbib]{revtex4}
\usepackage{}
\usepackage{amssymb}
\usepackage{bbm}
\usepackage{graphicx}
\usepackage{epsfig}
\usepackage{bm}
\usepackage{amsfonts}
\usepackage{epstopdf}
\usepackage{multirow}

\begin{document}

\title{A New Class of Parametrization for Dark Energy without Divergence }

\author{Chao-Jun Feng}
\email{fengcj@shnu.edu.cn} \affiliation{Shanghai United Center for Astrophysics (SUCA), \\ Shanghai Normal University,
    100 Guilin Road, Shanghai 200234, P.R.China}

\author{Xian-Yong Shen}
 \affiliation{Shanghai United Center for Astrophysics (SUCA), \\ Shanghai Normal University,
    100 Guilin Road, Shanghai 200234, P.R.China}

\author{Ping Li}
 \affiliation{Shanghai United Center for Astrophysics (SUCA), \\ Shanghai Normal University,
    100 Guilin Road, Shanghai 200234, P.R.China}
    
\author{Xin-Zhou Li}
\email{kychz@shnu.edu.cn} \affiliation{Shanghai United Center for Astrophysics (SUCA),  \\ Shanghai Normal University,
    100 Guilin Road, Shanghai 200234, P.R.China}

\begin{abstract}
In this paper, we propose a new class of parametrization of the equation of state of dark energy. In contrast with the famous CPL parametrization, these new parametrization of the equation of state does not divergent during the evolution of the Universe even in the future. Also, we perform a observational constraint on two simplest dark energy models belonging to this new class of parametrization, by using the Markov Chain Monte Carlo (MCMC) method and the combined latest observational data from the type Ia supernova compilations including Union2(557), cosmic microwave background, and baryon acoustic oscillation.
\end{abstract}


\maketitle


\section{Introduction}\label{sec:intro}
The cosmic acceleration has been discovered since 1998~\cite{Riess:1998cb, Perlmutter:1998np}, but we are still in the dark about the nature of this mystery. The cosmological constant proposed by Einstein to build a static Universe solution is the simplest dark energy model to explain the acceleration, but it suffers from the fine-tuning and coincidence problems. On the other side, current observations do not exclude definitely the time dependent evolution of dark energy, then many alternative dark energy models were proposed in the literatures, for a recent review, see~\cite{Li:2011sd, Bamba:2012cp}.

The evolution of dark energy is determined by its equation of state (EoS), which can be obtained in a particular model or constructed directly from observational data. In the later case, we usually make some assumptions on the form of the EoS as a parametrization for the dark energy model, then extract the information from observations. A most wildly used parametrization is called the Chevallier-Polarski-Linder (CPL) model~\cite{Chevallier:2000qy, Linder:2002et} with the form of EoS as
\begin{equation}\label{CPL}
  w(z) = w_0 + w_1 \frac{z}{1+z} \,,
\end{equation}
with redshift $z$. Here, $w_0$ is the present value of the EoS and $w_1$ is the derivative with respect to the scale factor. Comparing with the linear form $w(z) = w_0 + w_1 z$, the CPL parametrization is well behaved and bounded at high redshifts. This is the direct motivation of proposing such a  form (\ref{CPL}). However, the CPL form also suffers from the divergence problem when $z$ approaches $-1$, which corresponds to future evolution of the EoS. This is undoubtedly a nonphysical feature, and thus it prevents the CPL parametrization from genuinely covering the scalar field models as well as other theoretical models. Recently, a logarithm form $w(z) = w_0 + w_1\left[\ln(2+z)(1+z)^{-1} - \ln2\right]$ and an oscillating form  $w(z) = w_0 + w_1\left[\sin(1+z)(1+z)^{-1} - \sin1\right]$ are proposed in Ref.~\cite{Ma:2011nc, Li:2011dr, Li:2012vn} to overcome the shortcoming of the CPL model.

To extend parametrization of dark energy to redshifts $z \rightarrow -1$, we suggest two new models
 \begin{equation}\label{GCPLA}
 \text{Model I: } \qquad w(z) = w_0 + w_1 \frac{z}{1+z^2} \,,
\end{equation}
and
\begin{equation}\label{GCPLB}
 \text{Model II:} \qquad w(z) = w_0 + w_1 \frac{z^2}{1+z^2} \,,
\end{equation}
which are free of divergence in the whole range of the redshift $z\in[-1,\infty)$. Clearly, Model I reduce to the linear form when $z\ll 1$ as the CPL form, while Model II will be not. Actually, one can even generalize the form to a more general case while keeps the advantage of divergence-free, as long as some conditions are satisfied. We will make some discussions about it in the conclusion section, and one shall see that the forms (\ref{GCPLA}) and (\ref{GCPLB}) are the most simplest ones of this class of parametrization.

We perform a global data fitting analysis on these two divergence-free parametrization of dark energy models and constrain the parameters from the current observational data, including the full CMB power spectrum  from seven-year WMAP data (WMAP7)~\cite{Komatsu:2010fb}, the baryon acoustic oscillations (BAO) data from SDSS DR7~\cite{Percival:2009xn} and Union2 SNIa with $557$ samples~\cite{Amanullah:2010vv}. When doing the analysis, we also take the perturbation of dark energy into account, because the perturbation plays a important role when one confronts the dynamical dark energy models with the current observations~\cite{Zhao:2005vj, Li:2010hm, Hwang:2009zj, Feng:2009ag}. For instance, when the dark energy dominates the Universe at later time in a spatial flat model, the gravitational potential due to the perturbation of dark energy becomes time-dependent, thus there will be a significant imprint on the large-scale CMB power spectra by the integrated Sachs-Wolfe (ISW) effect~\cite{Sachs:1967er, Bean:2003fb, Schaefer:2008qs, Wang:2010zzj, Wang:2010cg, Sapone:2009mb}. We also make some analysis on this effect in this paper. We  added two new parameters $w_0, w_1$ and modified the public available \textbf{CosmoMC} package \cite{Lewis:2002ah} to satisfy  these two models. During the calculation, we run $32$ independent chains for each case, and we make sure the convergence of the chains by typically getting $\mathcal{R}-1$ to be less than $0.01$, see Ref.~\cite{Lewis:2002ah, Feng:2012gr, Ao:2011kc, Feng:2009zf, Liu:2008vy} and the readme file on the website of \textbf{CosmoMC}.

This paper is organized as follows: In Sec.\ref{sec:bpe}, we will study the background and perturbation evolutions  for both models. In Sec.\ref{sec:fitting}, we perform a global fitting to the current observational data by using the Markov Chain Monte Carlo (MCMC) method. Some analysis on the ISW effect of there two models are given in Sec.\ref{sec:isw}. In the last section, we will give some conclusions and make some discussions on how to generalize these models without losing the advantage of divergence-free.

\section{Background and Perturbation Evolution }\label{sec:bpe}

As we mentioned before, the new parametrizations of dark energy (\ref{GCPLA}) and (\ref{GCPLB}) are free of divergence in the whole range of the redshift $z\in[-1,\infty)$, especially when $z \rightarrow -1$. For clearness, we list the limiting values of $w(z)$ when $z\rightarrow \infty, -1, 0$ in Tab.\ref{table:limit} for the CPL model, Model I and Model II.
\begin{table}[h]
\centering
  \begin{tabular}{l|c|c|c|c}
  \hline
  \hline
  Model &$w(z)$& $ z = 0 $  & $z\rightarrow \infty$ & $z\rightarrow -1$\\
  \hline
  \hline
  CPL      &$w(z) = w_0 + w_1 z/(1+z)$     & \multirow{3}*{$w_0$}  & $w_0 + w_1$ & $\infty$ \\
  \cline{1-2}
  \cline{4-5}
  Model I  &$w(z) = w_0 + w_1 z/(1+z^2)$   &                       & $w_0$       & $w_0 - \frac{w_1}{2}$ \\
  \cline{1-2}
  \cline{4-5}
  Model II &$w(z) = w_0 + w_1 z^2/(1+z^2)$ &                       & $w_0 + w_1$ & $w_0 + \frac{w_1}{2}$ \\
  \hline
  \hline
  \end{tabular}
  \caption{\label{table:limit} The limiting values of the EoS of CPL, Model I and Model II. }
\end{table}
At low shifts, the Model I form will reduce to the linear form $w(z) \approx w_0 + w_1 z$ as the CPL model, while Model II will reduce to the quadratic form, i.e. $w(z) \approx w_0 + w_1 z^2$, and both of them are free of divergence. From the continuity equation of the dark energy with EoS (\ref{GCPLA}) and (\ref{GCPLB}), one can obtain the its energy density, and then the Friedmann equation would be
\begin{equation}\label{back}
  E^2 \equiv \frac{H^2}{H_0^2} = \Omega_M(1+z)^3 + \Omega_r(1+z)^4 + (1-\Omega_M-\Omega_r)\bigg[(1+z)^{3(1+ w_0)} \cdot f(z)\bigg] \,,
\end{equation}
for a spacial flat Universe. Here we have defined  $\Omega_i \equiv \rho_i / \rho_c, (i = b, m , r)$ and $\Omega_M = \Omega_b + \Omega_m$. The functions $f(z)_{\pm}$ are given by 
\begin{equation}
  f_{\pm}(z) = \exp\bigg[ \pm \frac{3w_1}{2} \arctan(z)\bigg](1+z)^{\mp \frac{3w_1}{2} }\left(1+z^2\right)^{\frac{3w_1}{4}}  \,,
\end{equation}
where $f_+$ and $f_-$ correspond to Model I and II, respectively. 

As we known, for dynamical dark energy models, namely, the EoS of dark energy is not constantly equal to $-1$, it is crucial to include dark energy perturbations~\cite{Zhao:2005vj, Li:2010hm, Hwang:2009zj, Feng:2009ag}. Working in the conformal Newtonian gauge with the perturbed metric as the following
\begin{equation}
  ds^2 = a(\eta)^2\bigg[ -(1+2\psi)d\eta^2 + (1-2\phi)\delta_{ij}dx^idx^j \bigg] \,,
\end{equation}
one can obtain the perturbed equations of dark energy as~\cite{Ma:1995ey,Bardeen:1980kt,Kodama:1985bj}
\begin{eqnarray}
  \delta' &=& -(1+w)(\theta - 3\phi') - 3\mathcal{H}(c_s^2-w)\delta \,, \label{perturb1}\\
  \theta' &=& -\mathcal{H}(1-3w)\theta - \frac{w'}{1+w}\theta + \frac{c_s^2}{1+w} k^2 \delta + k^2\psi \label{perturb2}\,,
\end{eqnarray}
where prime denotes the derivatives with respect to conformal time $\eta$, $\mathcal{H}= a'/a $ is the conformal Hubble parameter. Here $\delta$ and $\theta$ are the density and velocity perturbations respectively. The square sound speed $c_s^2$ is defined as $c_s^2 = \delta p/ \delta\rho$. It is clear that for the quintessence-like (or phantom-like) models with $w\ge-1$ (or $w\le-1$), the perturbation equations (\ref{perturb1}) and (\ref{perturb2}) are well-defined. In other words, as long as the EoS does not cross the cosmological boundary, the perturbation equations will not divergence. However, when the cross happens in the dark energy models such as CPL, Model I and II, one should take care when $w = -1$. This is so-called the divergence problem for perturbations of dark energy. A very useful practical solution to this problem is to introduce a small positive constant $\epsilon$, then the whole range of the allowed values of EoS is divided into three parts: $w>-1+\epsilon$, $w<-1-\epsilon$ and $-1-\epsilon\le w\le-1+\epsilon$, see Ref.~\cite{Zhao:2005vj}. For the first two regions, the perturbation equations are well-defined without divergence. To avoid the divergence in the last region, we take the approach that match the perturbations in this region to the first two regions at the boundary and set $\delta' = \theta' = 0$ in this region. In our numerical calculations, we limit the range to be $\epsilon < 10^{-7}$, suggested in Ref.~\cite{Wang:2011km}.

\section{Global Fitting Procedure and Results}\label{sec:fitting}

In this section, we will use the Markov Chain Monte Carlo (MCMC) method to make a global fitting on the cosmological parameters in Model I and II. We have modified the public available \textbf{CosmoMC} package \cite{Lewis:2002ah} to satisfy Model I and II, in which we have added two new parameters $w_0$ and $w_1$ and  the perturbation equations for dark energy. Our most general parameter space vector is
\begin{equation}\label{pv}
  \mathcal{P} \equiv \bigg\{ \Omega_b h^2, \Omega_c h^2, \Theta, \tau, w_0, w_1, n_s, \log[10^{10}A_s] \bigg\} \,,
\end{equation}
where $\Omega_b h^2$ and  $\Omega_c h^2$ are the physical baryon and cold dark matter densities, $\Theta$ is the ratio (multiplied by $100$) of the sound horizon to the angular diameter distance at decoupling, $\tau$ is the optical depth to re-ionization, $w_0$ and $w_1$ are the parameters of dark energy EoS given by Eqs.(\ref{GCPLA}) and (\ref{GCPLB}), $n_s$ is the scalar spectral index and $A_s$ is defined as the amplitude of the initial power spectrum. The pivot scale of the initial scalar power spectrum we used here is $k_{s0}=0.05$Mpc$^{-1}$ and we have assumed purely adiabatic initial conditions. In this paper, we will use the WMAP seven-year data~\cite{Komatsu:2010fb}, the BAO data~\cite{Percival:2009xn} and also the current supernova "Union2" data with $557$ simples~\cite{Amanullah:2010vv}. All the calculations are performed on the Astrophysical Beowulf Cluster (\textbf{AstroBC}), which is a parallel computing network system designed, set up and tested by ourselves, and it is much fast and stable than a single personal computer.

In order to determine the best value of parameters (with $1\sigma$  error at least ) in Model I and II, we
will use the maximum likelihood method and  take the total likelihood function $\mathcal{L} = e^{-\chi^2/2}$ as the products of these separate likelihood functions of each data set, thus we get
\begin{equation}\label{tot chi2}
    \chi^2 = \chi^2_{CMB}  +\chi^2_{BAO} + \chi^2_{SN}\,.
\end{equation}
Then, one can get the best fitting values of parameters by  minimizing $\chi^2$.

Our  global fitting results are presented in Tab.\ref{table:best}, in which we list the best fit  values with $1\sigma, 2\sigma$ errors for each parameter in the parameter space vector (\ref{pv}), as well as the constraint for the age of the Universe. For comparison, we also present the fitting results without dark energy perturbations in  Tab.\ref{table:best}, and we find  that the results with perturbations show a better fit to that without perturbations by comparing the minimal values of $\chi^2$. In our analysis, we have fixed the sound speed of dark energy in the rest frame $\tilde c_s^2$, which related to $c_s^2$ in a general frame as \cite{Bean:2003fb, Kodama:1985bj}
\begin{equation}
  c_s^2 = \tilde c_s^2 + 3 \mathcal{H}(1+w)(\tilde c_s^2 - c_a^2) \frac{\theta}{k^2} \delta^{-1}\,.
\end{equation}
To derive the above relation, we have used the useful relation between the gauge invariant, rest frame density perturbation $\tilde \delta$ and the density and velocity perturbations in a general frame, $\delta$ and $\theta$
\begin{equation}
  \tilde\delta = \delta + 3\mathcal{H}(1+w)\frac{\theta}{k^2}\,,
\end{equation}
and the definition of the adiabatic sound speed
\begin{equation}\label{acs}
  c_a^2 \equiv \frac{p'}{\rho'} = w - \frac{w'}{3\mathcal{H}(1+w)} \,,
\end{equation}
see Ref.\cite{Bean:2003fb, Kodama:1985bj}. Since it has been shown that the constraints on the dark  energy sound speed  in the rest frame $\tilde c_s^2$ in dynamical dark energy models are still not accurate enough~\cite{Xia:2007km}, then one can treats the dark energy model as a effective scalar-field model and sets $\tilde c_s^2$ to be unit~\cite{Li:2011dr}. The results would not be affected by this treatment if $\tilde c_s^2$ is considered as a free parameter. In this paper, we have taken the full CMB temperature and polarization power spectra instead of just taking the WMAP distance prior, including $R, l_A$ and $z_*$, which encoding the background distance information and can be used to investigate dark energy models with greatly simplified numerical calculations. However, it was found that this distance prior is somewhat cosmological model dependent, thus, one would loss some CMB information by using it to fit the parameters~\cite{Komatsu:2010fb}. For instance, as we known, if there was a time-dependent gravitational potential at later times, it will contribute to the ISW effect. Therefore, our results  would be more reliable than that obtained by using just the distance prior.
\begin{table}[h]
\centering
  \begin{tabular}{c|c|c|c|c}
  \hline
  \hline
  \multirow{2}{*}{ Model $\&$  Parameter } & \multicolumn{2}{c|} {Model I}  & \multicolumn{2}{c|} {Model II} \\
  \cline{2-5}
  & w/. perturb.  & w/o. perturb.& w/. perturb.& w/o. perturb. \\
  \hline
  \multicolumn{1}{c|}{Data} & \multicolumn{4}{c}{ CMB $+$ BAO $+$ SN  } \\
  \hline
  \hline
  $\Omega_b h^2$      & $0.0223^{+0.0016+0.0021}_{-0.0014-0.0019}$  & $0.0222^{+0.0015+0.0019}_{-0.0015-0.0020}$
                      & $0.0225^{+0.0015+0.0018}_{-0.0017-0.0021}$  & $0.0221^{+0.0016+0.0020}_{-0.0015-0.0018}$ \\
  \hline
  $\Omega_{DM} h^2$   & $0.1094^{+0.0126+0.0155}_{-0.0133-0.0163}$  & $0.1101^{+0.0133+0.0167}_{-0.0126-0.0148}$
                      & $0.1107^{+0.0118+0.0161}_{-0.0141-0.0172}$  & $0.1101^{+0.0139+0.0178}_{-0.0116-0.0136}$ \\
  \hline
  $\Theta$            & $1.0387^{+0.0074+0.0094}_{-0.0071-0.0098}$  & $1.0372^{+0.0086+0.0104}_{-0.0066-0.0081}$
                      & $1.0391^{+0.0069+0.0088}_{-0.0078-0.0097}$  & $1.0382^{+0.0076+0.0092}_{-0.0072-0.0094}$   \\
  \hline
  $\tau$              & $0.0882^{+0.0484+0.0574}_{-0.0375-0.0469}$  & $0.0814^{+0.0526+0.0679}_{-0.0319-0.0409}$
                      & $0.0867^{+0.0475+0.0619}_{-0.0379-0.0453}$  & $0.0853^{+0.0449+0.0614}_{-0.0333-0.0473}$   \\
  \hline
  $w_0$               & $-1.0148^{+0.5183+0.6482}_{-0.5907-0.7247}$ & $-1.0153^{+0.6181+0.7664}_{-0.5576-0.7103}$
                      & $-1.0214^{+0.3093+0.4287}_{-0.2846-0.3373}$ & $-0.9463^{+0.3304+0.4165}_{-0.3628-0.4743}$  \\
  \hline
  $w_1$               & $-0.0155^{+2.1257+2.6973}_{-2.0833-2.7378}$ & $-0.1146^{+2.0636+2.6046}_{-2.4806-3.3328}$
                      & $-0.0113^{+0.9683+1.1182}_{-2.5469-3.9020}$ & $-0.4783^{+1.5773+1.7560}_{-3.5730-4.9333}$  \\
  \hline
  $n_s$               & $0.9657^{+0.0373+0.0496}_{-0.0373-0.0448}$  & $0.9579^{+0.0373+0.0458}_{-0.0329-0.0432}$
                      & $0.9632^{+0.0410+0.0491}_{-0.0351-0.0432}$  & $0.9565^{+0.0404+0.0473}_{-0.0317-0.0407}$   \\
  \hline
  $\log[10^{10} A_s]$ & $3.0716^{+0.1066+0.1269}_{-0.0960-0.1276}$  & $3.0532^{+0.1187+0.1494}_{-0.0766-0.1072}$
                      & $3.0695^{+0.1075+0.1374}_{-0.1062-0.1177}$  & $3.0578^{+0.1149+0.1402}_{-0.0830-0.1003}$   \\
  \hline
  Age/Gyr             & $13.7671^{+0.3669+0.4684}_{-0.3569-0.4282}$ & $13.8014^{+0.3596+0.4287}_{-0.3608-0.4255}$
                      & $13.7338^{+0.4803+0.5680}_{-0.3471-0.4094}$ & $13.7393^{+0.4819+0.5867}_{-0.3627-0.4464}$   \\
  \hline
  \hline
  $\chi^2_{min}$      & $8000.764$ & $8001.348$
                      & $8000.958$ & $8001.184$ \\
  \hline
  \hline
  \end{tabular}
  \caption{\label{table:best} The fitting results of the parameters with $1\sigma, 2\sigma$ regions in Model I and Model II with and without perturbations from the combination of WMAP, BAO and SN. }
\end{table}

\begin{figure}[h]
\begin{center}
\includegraphics[width=1.0\textwidth,angle=0]{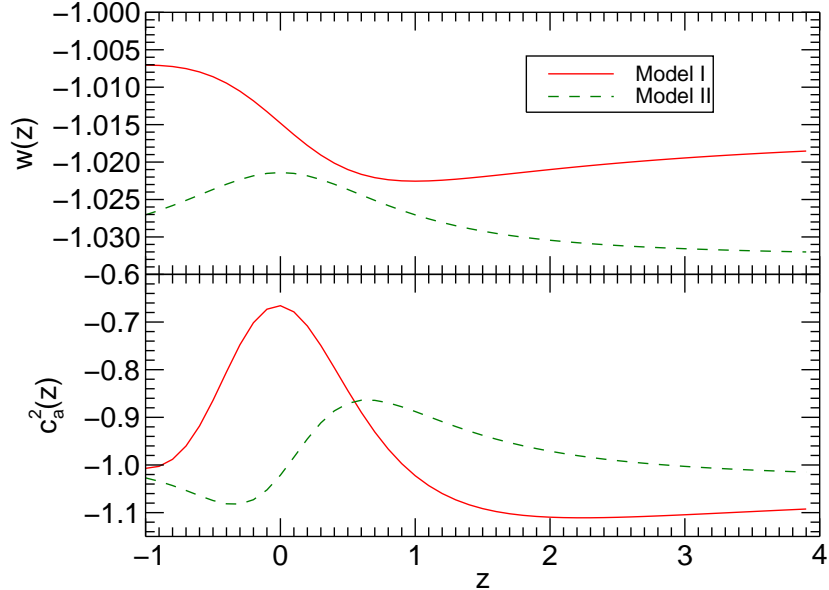}
\caption{\label{fig::wz} The evolution of EoS $w(z)$ and the adiabatic square sound speed $c_a^2$ in Model I and II with their best-fit values of $w_0$ and $w_1$. }
\end{center}
\end{figure}

From Tab.\ref{table:best}, one can see that in both models, the best-fit values of the scalar spectral index $n_s$ are smaller than $1$. That is to say, the scalar spectrum is "red" tiled. The difference between Model I and II is shown in Fig.\ref{fig::wz}, in which we have plotted the evolution of the EoS  and the adiabatic square sound speed for both models with their best-fit values of $w_0$ and $w_1$. From Fig.\ref{fig::wz}, we find that the best-fit dark energy model is a phantom model, whose $w(z)$ is always smaller than $-1$, and the evolution of $w(z)$ is quite different between these two models. However, the difference of the present values of $w(z)$ is smaller than one percent. These results imply that though the dynamical dark energy models are mildly favored, the current data cannot distinguish them effectively.

\begin{figure}[h]
\begin{center}
\includegraphics[width=1.0\textwidth,angle=0]{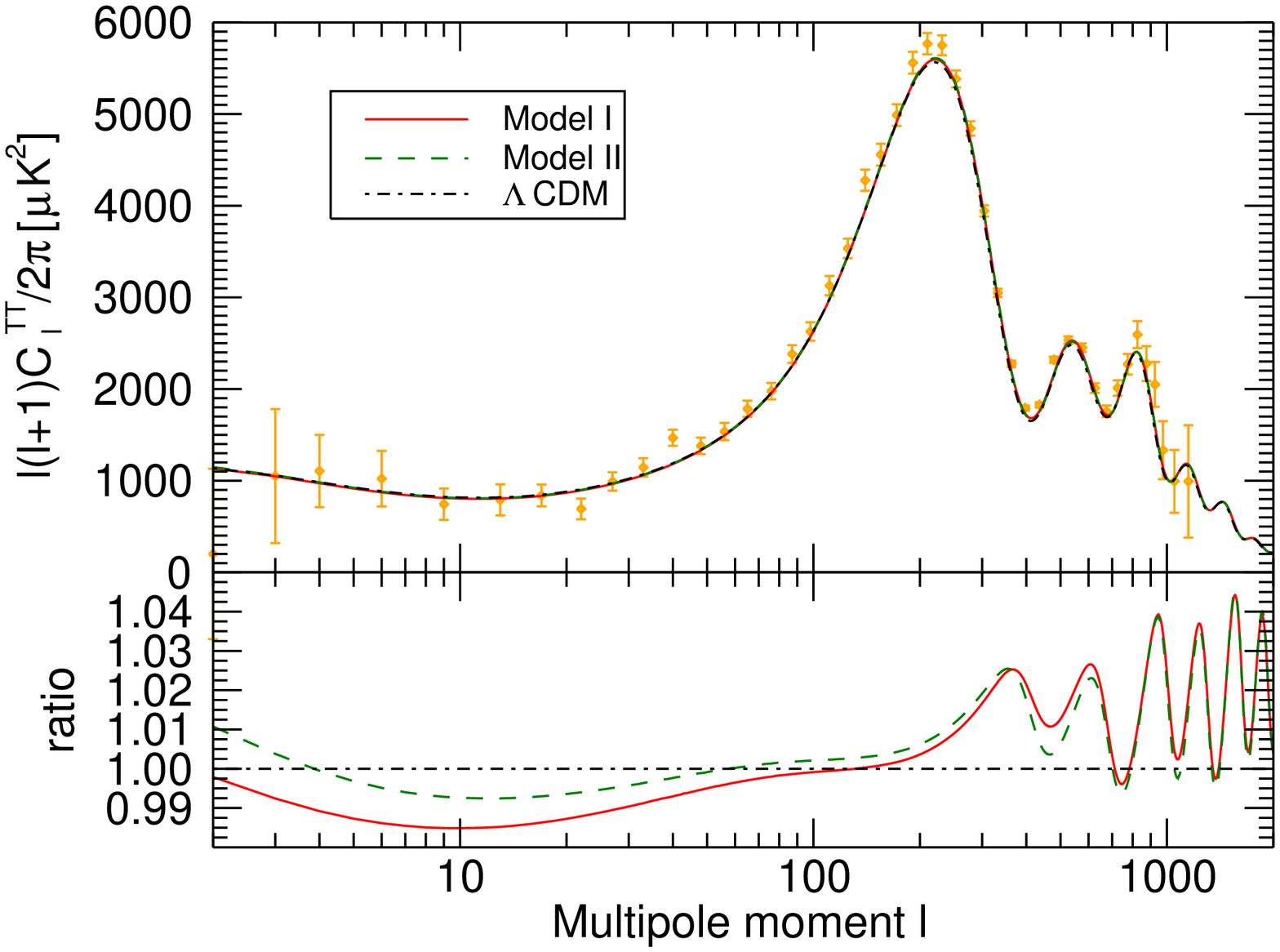}
\caption{\label{fig::ctt} The CMB  $C_l^{TT}$ power spectrum v.s. multipole moment $l$, where the yellow dots with error bars denote the observational data with their corresponding uncertainties from WMAP 7-year results, the red solid line, red dashed and black dotted-dashed curves are for Model I, II and $\Lambda$CDM model with their best-fit values from WMAP+BAO+SN joint constraint. }
\end{center}
\end{figure}

\begin{figure}[h]
\begin{center}
\includegraphics[width=1.0\textwidth,angle=0]{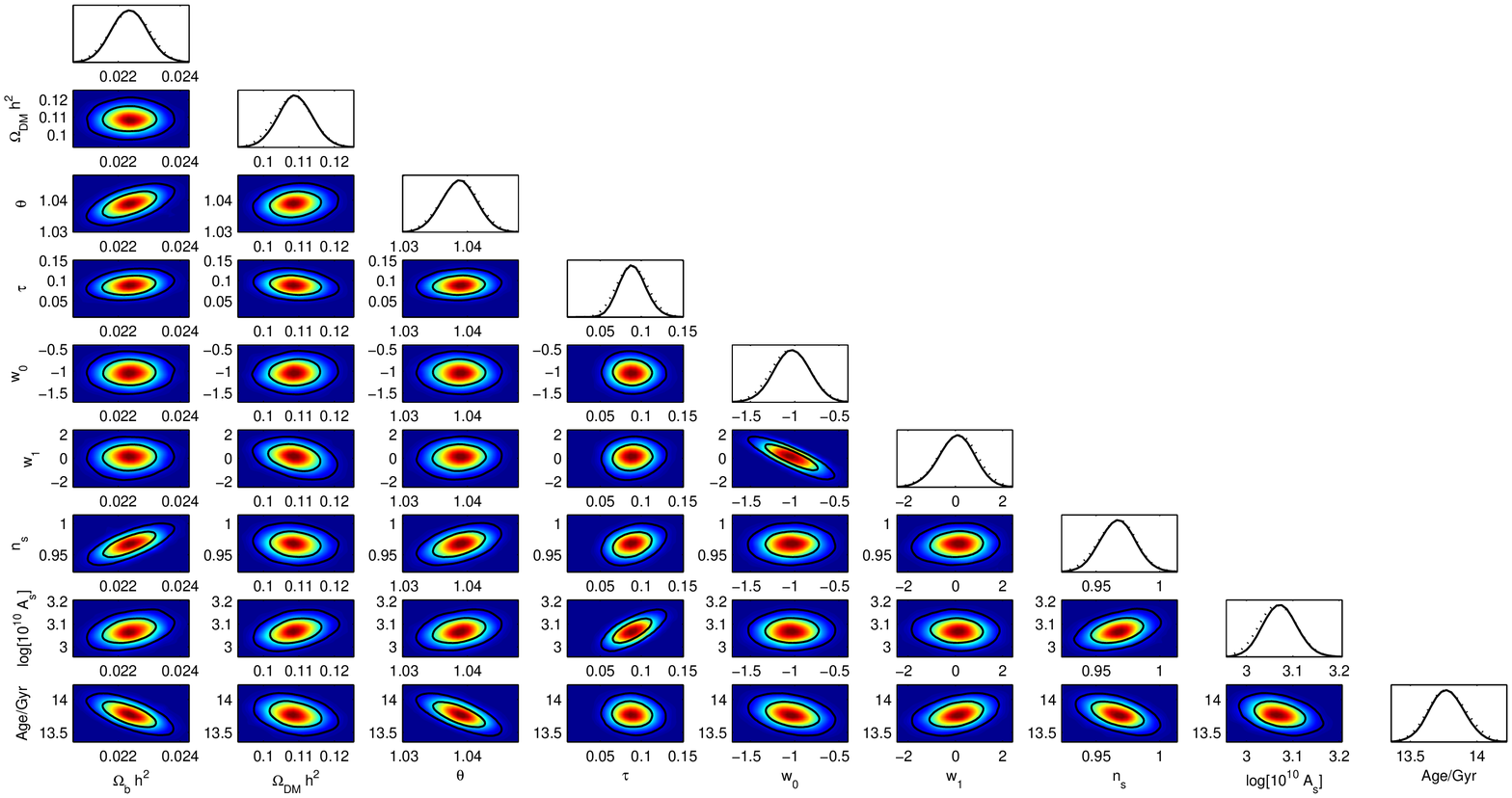}
\caption{\label{fig::Model1} The $1$-D constraints on each parameters in  $\mathcal{P}\equiv \left\{ \Omega_b h^2, \Omega_c h^2, \Theta, \tau, w_0, w_1, n_s, \log[10^{10}A_s] \right\}$ with the age of Universe, and $2$-D contours on these parameters with $1\sigma, 2\sigma$ confidence level each other in Model I.  Here the dotted curves in each $1$-D plots are the mean likelihood of the samples, while the solid curves are the marginalized probabilities for each parameters. }
\end{center}
\end{figure}

\begin{figure}[h]
\begin{center}
\includegraphics[width=1.0\textwidth,angle=0]{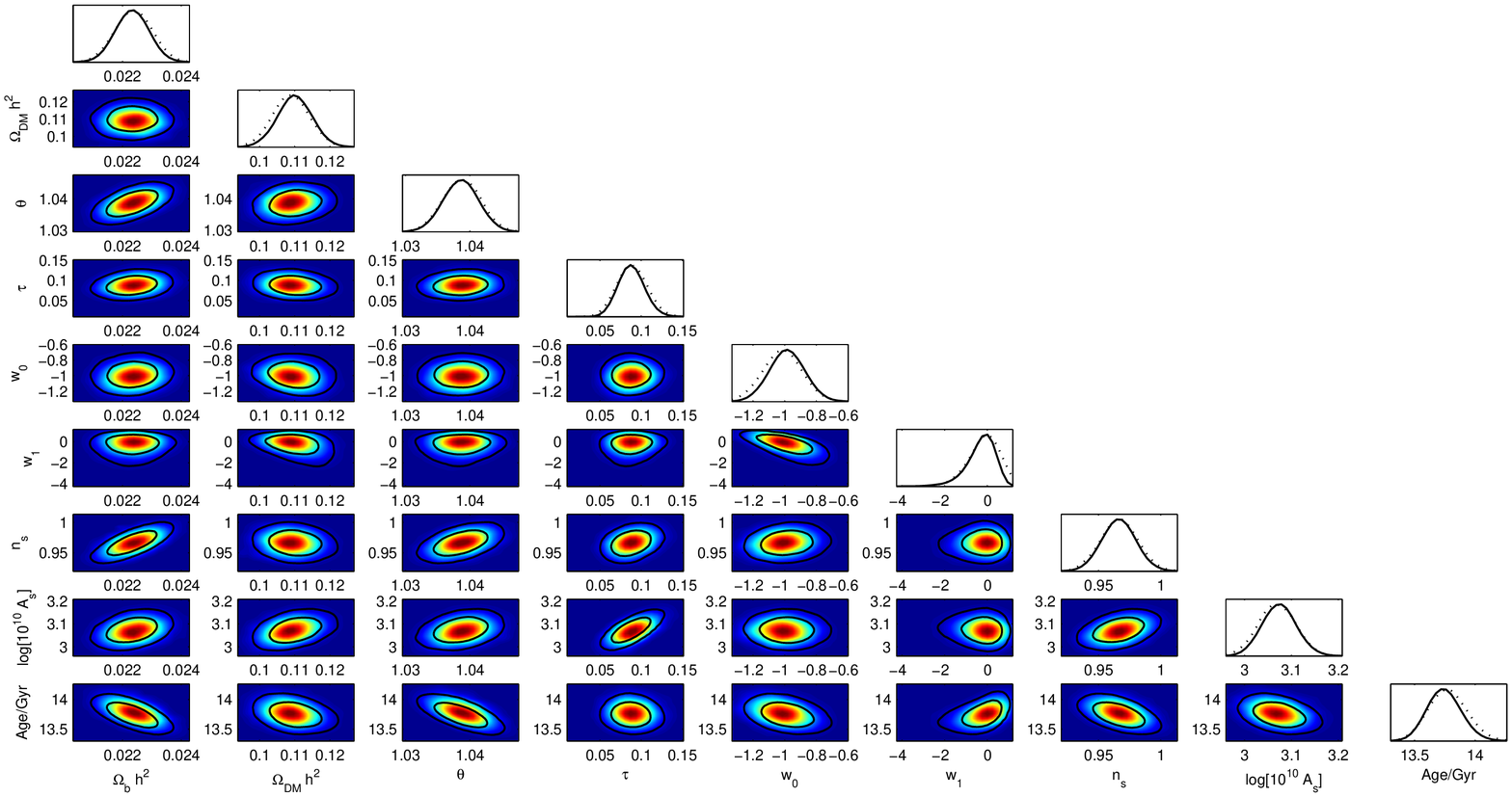}
\caption{\label{fig::Model2} The $1$-D constraints on each parameters in  $\mathcal{P}\equiv \left\{ \Omega_b h^2, \Omega_c h^2, \Theta, \tau, w_0, w_1, n_s, \log[10^{10}A_s] \right\}$ with the age of Universe, and $2$-D contours on these parameters with $1\sigma, 2\sigma$ confidence level each other in Model II.  Here the dotted curves in each $1$-D plots are the mean likelihood of the samples, while the solid curves are the marginalized probabilities for each parameters. }
\end{center}
\end{figure}

In Fig.~\ref{fig::ctt}, we show the CMB temperature power spectrum for Model I and II with their best-fit parameters listed in Tab.~\ref{table:best}. The spectrum for $\Lambda$CDM is also plotted for comparison. We find that the most significant discrepancies between Model I, II and $\Lambda$CDM come from the low $l$s, or large angular scales ($\sim 180^{\circ}/l$). The contribution to the secondary CMB anisotropy is mainly contributed from the late ISW effect, see the discuss in next section. Both cures of Model I and II tend to be plat on large scales. In the Fig.\ref{fig::Model1} and Fig.\ref{fig::Model2}, we also show the one dimensional probability distribution of each parameters and  also the two dimensional contour plots between each other in the Model I and II from combination of WMAP, SN and BAO.

\section{The ISW Effect}\label{sec:isw}
The ISW effect is due to the existence of time-dependent gravitational potentials along the line of sight from the epoch of last scattering up to now~\cite{Sachs:1967er, Hu:1994jd, Hu:1995sv}.  There is a net change in the energy of a CMB photon as it passes through these evolving gravitational potential wells, and the sum of these contributions is called the ISW effect.  The ISW contribution to the CMB power in the $l$th multipole is given by
\begin{equation}\label{iswcl}
  \frac{2l+1}{4\pi}C_l^{ISW} = \frac{1}{2\pi^2} \int_0^{\infty} \frac{dk}{k}k^3 \frac{|\Theta_l^{ISW}(\eta_0, k)|^2}{2l+1} \,,
\end{equation}
where $\Theta_l^{ISW}$ is the multipole decomposition of the fractional fluctuation due to the ISW effect
\begin{equation}\label{isw}
  \frac{\Theta_l^{ISW}(\eta_0,k)}{2l+1} = 2 \int_{\eta_*}^{\eta_0} \psi'(\eta,k) j_l[k(\eta_0 - \eta)] d\eta \,.
\end{equation}
Here, $\eta_0$ and $\eta_*$ denote the present and the last scattering conformal times respectively, and $j_l$ are the spherical Bessel functions.  As we known, the perturbation equation of $\psi = 4\pi G(\rho+p)^{1/2} u $ is given by
\begin{equation}\label{uequ}
  u'' - c_s^2 \nabla^2 u- \frac{\theta''}{\theta} u = 0 \,,
\end{equation}
where $\theta \equiv (1+ w)^{-1/2} a^{-1}$. The adiabatic square sound speed  in a dark energy dominated Universe is given by Eq.~(\ref{acs}), and one can rewritten $c_a^2$ in terms of redshift
\begin{equation}
   c_s^2  = w + \frac{1+z}{3(1+w)} \frac{dw}{dz} \,,
\end{equation}
where $w$ takes the form of (\ref{GCPLA}) and (\ref{GCPLB}). And thus, we obtain
 \begin{equation}\label{csGCPLA}
 \text{Model I: } \qquad c_a^2(z) = w_0 + \frac{z w_1}{1+z^2}
 + \frac{(1-z)(1+z)^2 w_1}{3(1+z^2)(1+z^2 + (1+z^2)w_0 + zw_1)}  \,,
\end{equation}
and
\begin{equation}\label{csGCPLB}
 \text{Model II:} \qquad c_a^2(z) = w_0 + \frac{z^2 w_1}{1+z^2}
 + \frac{2z(1+z)w_1}{3(1+z^2)(1+z^2 + (1+z^2)w_0 + z^2 w_1)}\,.
\end{equation}
The evolution of $c_a^2$ is shown in Fig.~\ref{fig::wz} in Model I and II.

It is usual to divide the ISW effect into early and late parts. The early ISW effect corresponds to the variation in the perturbation potentials which occurs in the time interval between matter-radiation equality and recombination, while the late ISW effect corresponds to the variations caused by the presence of dark energy. The expansion time scale at dark energy domination $\eta_{de} $ sets a critical wavelength corresponding to $k\eta_{de}=1$. Thus, for adiabatic perturbations, the late ISW effect in Eq.~(\ref{isw}) during the dark energy epoch $\eta\gg \eta_{de}$ for a given mode can be approximated by~\cite{Hu:1994jd, Hu:1995sv}
\begin{equation}
  \frac{\Theta_l^{ISW}(\eta_0,k)}{2l+1} \approx 2\frac{I_l}{k}\psi'(\eta_k, k) \,,
\end{equation}
where $\eta_k = \eta_0 - (l+1/2)/k$ is the conformal time when a given $k$-mode contributes maximally to the angle that this scale subtends on the sky, which obtained at the peak of the Bessel function $j_l$. Here the integral $I_l$ is given by
\begin{equation}
  I_l \equiv \int_0^\infty j_l(x)dx = \frac{\sqrt{\pi}}{2} \frac{\Gamma[(l+1)/2]}{\Gamma[(l+2)/2]} \,.
\end{equation}

Substituting the results above into Eq.~(\ref{iswcl}), we obtain the $C_l$ momenta of the late ISW effect
\begin{equation}\label{iswf}
  C_l^{ISW} \approx 2 \bigg( \frac{\Gamma[(l+1)/2]}{\Gamma[(l+2)/2]} \bigg)^2 \int_0^\infty \psi^{'2}(\eta_k,k) dk \,.
\end{equation}
Therefore, one can at least numerically obtain the solution of $\psi$ from Eq.~(\ref{uequ}) and then get $C_l$s. From Eq.~(\ref{iswf}), one can see that the main contribution of the late ISW effect mainly comes from the low momenta $l$, see Fig.~\ref{fig::ctt}.

\section{Discussion and Conclusion}
In this paper, we have proposed two new parametrizations for the EoS of dark energy, see Eqs.(\ref{GCPLA}) and (\ref{GCPLB}). Actually, these two models belong to the following general case
\begin{equation}\label{genwz}
  w(z) = w_0 + w_1 \frac{z^m}{1+z^{2n}} \,,
\end{equation}
where Model I corresponds to $(m=1, n=1)$ and Model II corresponds to $(m=2, n=1)$. From the above equation, one can see that as long as $n$ is a non-zero integer, $w(z)$ would be not divergent when $z\rightarrow -1$, and the limiting value is $w(z\rightarrow-1)= w_0 + (-1)^m w_1/2$. Furthermore, if $m\le 2n$, the EoS would be also not divergent when the redshift goes to infinite. The limiting values for $w(z)$ when $z = 0$ and $z\rightarrow\infty$ are $w(0) = w_0$ and $w(z\rightarrow\infty) = w_0 (m<2n), w_0 + w_1 (m=2n)$. In general, one can not get an analytic result for the function $f(z)$ in the background equation (\ref{back}), and usually  it is given by the following integration
\begin{equation}
  f(z) = \exp\left[ 3w_1\int_0^z \frac{\tilde z^{m} d\tilde z}{(1+\tilde z^{2n})(1+\tilde z)} \right]\,.
\end{equation}
It is interesting that in Model I, the EoS could be rewritten as
\begin{equation}
  w(z) = w_0 + \frac{ w_1 }{z+z^{-1}} \,,
\end{equation}
from which one can see that $w(z)\approx w_0 + w_1/z$ when the redshifts are large ($z\gg1$), while $w(z)\approx w_0 + w_1 z$ when the redshifts are small ($z\ll1$), in other words, the EoS of Model I has the symmetry of $z \leftrightarrow z^{-1}$.

In conclusion, we have performed a global fit study on two parametrization models for dark energy, named Model I and II. In contrast to the famous CPL parametrization, in which $w(z)$ is divergent when $z\rightarrow -1$, Model I and II overcome this  shortcoming and are divergence-free in the whole range of the redshifts $z\in[-1, \infty)$. We have not only obtained the best-fit values with $1\sigma$ and $2\sigma$ regions for both of Model I and II with and without dark energy perturbations, but also found that the differences of CMB temperature power spectrum $C_l^{TT}$ on large angular scales are significant between Model I, II and $\Lambda$CDM model. There  also exists a flat plateau on the large scales in both models. In view of this difference, on which the ISW effect plays an important role. The effects on the parametrizations of Model I and II made by the ISW data from the correlations of CMB and large scale structure will be studied in our further works. And the class of this parametrization (\ref{genwz}) is also worth investigating in future.

\acknowledgments
This work is supported by National Science Foundation of China grant Nos.~11105091 and~11047138, National Education Foundation of China grant  No.~2009312711004, Shanghai Natural Science Foundation, China grant No.~10ZR1422000,  and  Shanghai Special Education Foundation, No.~ssd10004.
\appendix

\end{document}